\documentclass{PoS}
\usepackage{graphicx}

\title{Form Factors and Dyson-Schwinger Equations}

\ShortTitle{Dyson-Schwinger equations}

\author{I.C.\ Clo\"{e}t \\
        Department of Physics, University of Washington, Seattle WA 98195, USA\\
        E-mail: \email{icloet@phys.washington.edu}
        }
        
\author{\speaker{C.D.\ Roberts} \\
        Physics Division, Argonne National Laboratory, Argonne IL 60439, USA\\
        E-mail: \email{cdroberts@anl.gov}}

\abstract{A synopsis exemplifying the employment of Dyson-Schwinger equations in the calculation and explanation of hadron electromagnetic form factors and related phenomena.  In particular the contribution: presents the pion form factor computed simultaneously at spacelike and timelike momenta; reports aspects of the evolution of the nucleon and $\Delta$ masses with current-quark mass and the correlation of their mass difference with that between scalar and axial-vector diquarks; describes an estimate of the $s$-quark content of a dressed $u$-quark and its impact on the nucleon's strangeness magnetic moment; and comments upon the domain within which a pseudoscalar meson cloud can materially contribute to hadron form factors.
}

\FullConference{LIGHT CONE 2008 Relativistic Nuclear and Particle Physics\\
		 July 7-11, 2008\\
		 Mulhouse, France}

\begin{document}

\section{Proem}
In part owing to the simplicity of the photon as a probe, an accurate description of electromagnetic form factors provides information on the distribution of a hadron's characterising properties (total- and angular-momentum, etc.) amongst its QCD constituents.  Since contemporary experiments employ $Q^2>M^2$; i.e., momentum transfers in excess of the hadron's mass, a veracious understanding of the body of extant data requires a Poincar\'e covariant description of the hadron.  In fact the challenge is greater.  Owing to the running of the dressed-quark mass \cite{Bhagwat:2003vw,Bhagwat:2006tu} and related phenomena \cite{Bhagwat:2004hn,Bhagwat:2004kj}, a quantum field theoretic treatment of hadron structure and reactions is generally necessary to provide genuine understanding in terms of QCD's elementary degrees of freedom.  It is important to appreciate that Poincar\'e covariance and the vector exchange nature of QCD guarantee the existence of nonzero quark orbital angular momentum in a hadron's rest-frame bound-state amplitude \cite{Bhagwat:2006xi,Cloet:2007pi}.

In QCD the quark-parton acquires a momentum-dependent mass function, which at infrared momenta is $\sim 100$-times larger than the current-quark mass.  The Dyson-Schwinger equations (DSEs) \cite{Roberts:1994dr,Roberts:2007jh}
explain that this effect owes primarily to a dense cloud of gluons that clothes a low-momentum quark \cite{Bhagwat:2007vx,Roberts:2007ji}.  This marked momentum-dependence of the dressed-quark mass function is one manifestation of dynamical chiral symmetry breaking (DCSB).  It entails that the Higgs mechanism is largely irrelevant to the bulk of normal matter in the universe.  Instead the single most important mass generating mechanism for light-quark hadrons is the strong interaction effect of DCSB; e.g., one can identify it as being responsible for roughly 98\% of a proton's mass. 

Understanding the relationship between parton properties on the light-front and the rest frame structure of hadrons is a longstanding challenge.  It is a problem because, e.g., DCSB, an established keystone of low-energy QCD, has not been realised in the light-front formulation.  The obstacle is the constraint $k^+:=k^0+k^3>0$ for massive quanta on the light front \cite{Brodsky:1991ir}.  It is therefore impossible to make zero momentum Fock states that contain particles and hence the vacuum is trivial.  Only the zero modes of light-front quantisation can dress the ground state but little progress has been made with understanding just how that might occur.  It is noteworthy that DCSB has a valid expression solely within a framework that manifestly supports the axial-vector Ward-Takahashi identities.  Absent this its corollaries can only be obtained by fine tuning model-dependent inputs.

\section{Form Factors}
An explanation of pion and nucleon structure and interactions is central to hadron physics because they are respectively the archetypes for mesons and baryons.  Elastic and transition form factors have long been recognised as a basic tool for elucidating bound state properties.  They can be studied from very low momentum transfer, the region of non-perturbative QCD, up to a region where perturbative QCD predictions can be tested.  Experimental and theoretical studies of nucleon electromagnetic form factors have made rapid and significant progress during the last several years, including new data in the time like region, and material gains have been made in studying the pion form factor.  Despite this, many questions remain unanswered, amongst them:
can one formulate an impulse-like approximation for hadron form factors and, if so, in terms of which degrees of freedom;
what role is played by pseudoscalar mesons in hadron electromagnetic structure and can one describe this in a quantitative, model-independent fashion;
and what is the nature of hadron form factors in the timelike region and their quantitative connection with the spacelike behaviour?
The current status is described in Refs.\,\cite{Arrington:2006zm,Perdrisat:2006hj}

\section{Pion}
While the pion might be viewed as an archetype for mesons it has some remarkable and peculiar features.  Indeed, as QCD's Goldstone mode its pointwise and global properties are influenced to an enormous degree by DCSB.  True understanding is impossible in an approach that does not possess a valid and well-defined chiral limit, and an expression of the axial-vector Ward-Takahashi identity.  This DSE identity relates the gap and Bethe-Salpeter equations.  The gap equation produces the dressed-quark mass, and it is through the dressed-quark mass function that the connection between current- and constituent-quarks is explained \cite{Roberts:2007ji}.  

\begin{figure}[t]
\begin{minipage}[t]{0.45\textwidth}
\leftline{\includegraphics[clip,width=0.7\textwidth,angle=-90]{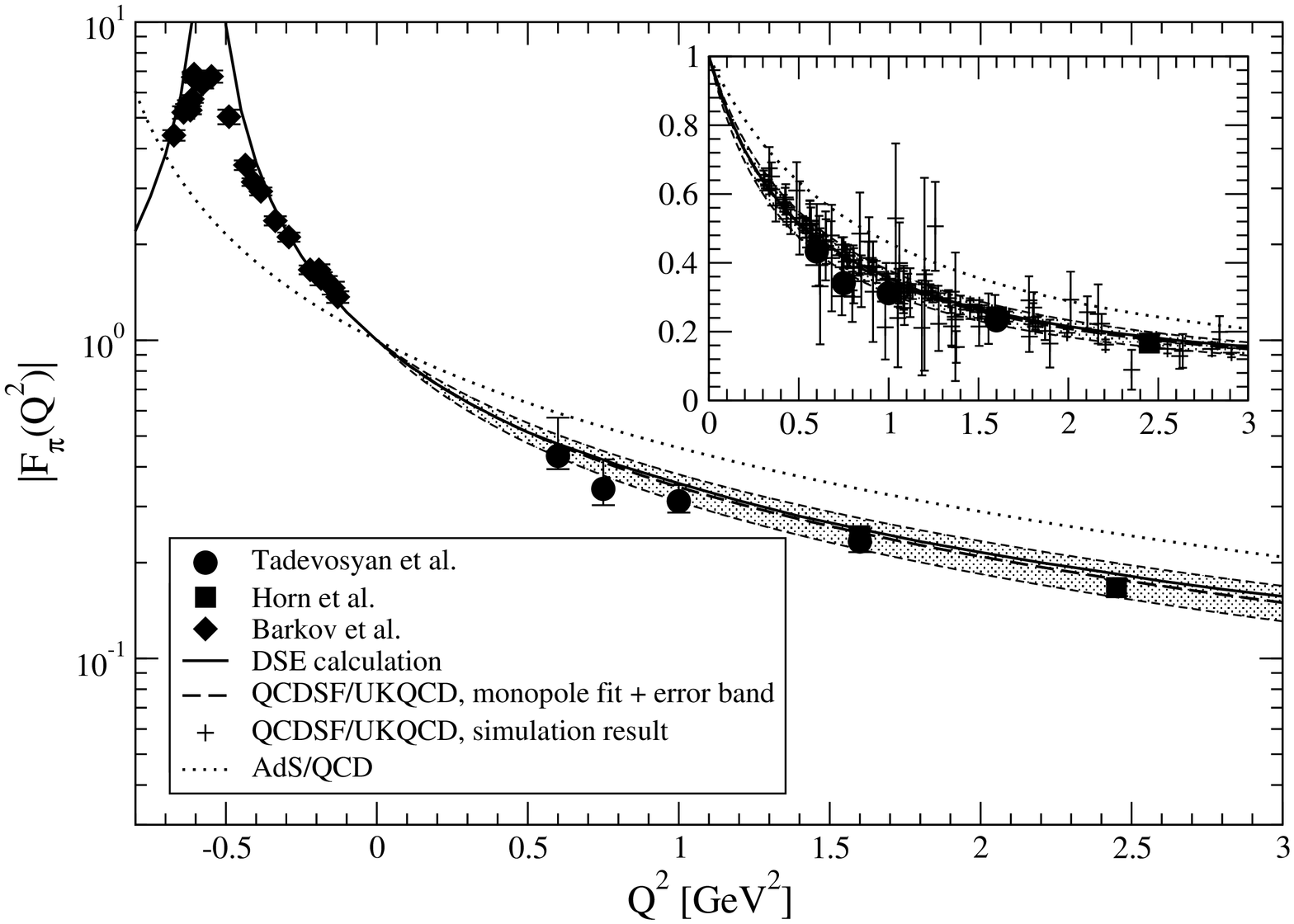}}
\end{minipage}
\begin{minipage}[t]{0.45\textwidth}
\rightline{\includegraphics[clip,width=0.70\textwidth,angle=-90]{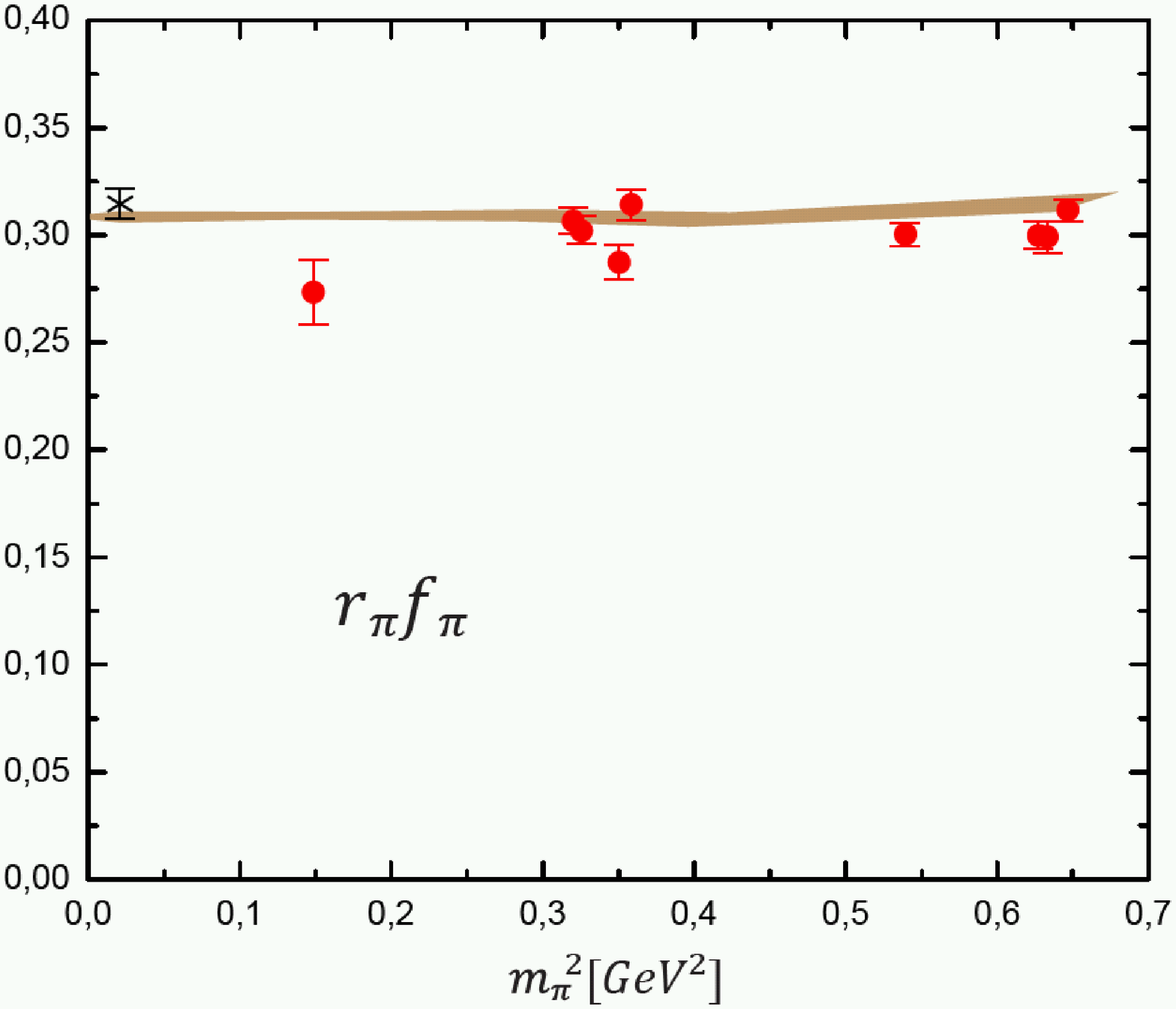}}
\end{minipage}\vspace*{3ex}
\caption{\label{AKFpitimelike} \underline{Left Panel}. \emph{Solid curve}: \emph{Ab-initio} DSE prediction of pion form factor.  The $\rho$-meson pole is generated dynamically.  No vector meson dominance assumption is made.  Depicted also are lattice results with a monopole fit \cite{Brommel:2006ww} (\emph{insert} and \emph{dashed curve}) and the result obtained from an AdS/QCD model \cite{Grigoryan:2008cc} with its parameter fitted to reproduce the pion's leptonic decay constant (\emph{dotted curve}). Data: \emph{diamonds}, Ref.\,\cite{Barkov:1985ac}; \emph{squares}, Ref.\,\cite{Horn:2006tm}; and \emph{circles}, Ref.\,\cite{Tadevosyan:2007yd}.  (Figure courtesy of A.~Krassnigg.) \underline{Right Panel}.  \emph{Band} -- DSE prediction for the current-quark mass dependence of the dimensionless product $r_\pi f_\pi$.  The band's width delineates the response to $\pm 20$\% variations in the interaction's range. \emph{Cross} -- experimental value: $0.315\pm0.005$.  \emph{Filled circles} -- Lattice-QCD result as determined \cite{James} from Ref.\,\cite{Brommel:2006ww}.  The AdS/QCD model predicts $r_\pi^2 f_\pi^2 = 9/[16 \pi^2] = (0.24)^2$. (Figure courtesy of G.~Eichmann.)}
\end{figure}

The existence of a sensible DSE truncation \cite{Munczek:1994zz,Bender:1996bb} has enabled proof of numerous exact results for pseudoscalar mesons \cite{Bhagwat:2006xi,Holl:2004fr,Holl:2005vu,Ivanov:1997yg,Ivanov:1998ms}.  They have been illustrated using a renormalisation-group-improved rainbow-ladder truncation, which also provides, e.g., a prediction of the electromagnetic pion form factor using an impulse approximation current that can systematically be improved \cite{Maris:2000sk}.  In building this current the basic degree of freedom is the dressed-quark.  The calculated form factor is depicted in Fig.\,\ref{AKFpitimelike} \cite{AKprivate}.

The illustrations employ a kernel of the gap and Bethe-Salpeter equations that is exact on the domain within which a perturbative calculation is valid.  Outside this domain it expresses a model for the long-range interaction between light-quarks, which is defined via a single parameter; viz., $\omega$: $r_a=1/\omega$ specifies the interaction's range and thereby a confinement length-scale.\footnote{NB.\ The potential between infinitely-heavy quarks measured in numerical simulations of quenched lattice-regularised QCD is not relevant to the question of light-quark confinement.  One cannot speak of a quantum mechanical potential between light-quarks because particle creation and annihilation effects are essentially nonperturbative.} The curve in Fig.\,\ref{AKFpitimelike} was obtained with $\omega$ fitted to reproduce the pion's leptonic decay constant.  This calculation is part of a programme that uses the interplay between experiment and theory as a means by which to map out the infrared behaviour of QCD's \emph{universal} $\beta$-function.  It is important to appreciate that while this function may depend on the scheme chosen to renormalise the quantum field theory, it is unique within a given scheme.

It was recently established \cite{Eichmann:2008ae} that in connection with light-quark systems, and those of the physical qualities of the pseudoscalar and vector meson bound states they constitute which are not tightly constrained by symmetries, the rainbow-ladder truncation of QCD's DSEs should produce results that, when measured in units of mass, are uniformly $\approx 35$\% too large.  The systematic implementation of corrections then shifts calculated results so that reliable predictions and agreement with experiment can subsequently be expected.  One can arrive in this way at a veracious understanding of light-quark observables.  It was also verified that the renormalisation-group-improved rainbow-ladder kernel predicts values for such observables that are insensitive to $\pm$ 20\% variations in $\omega$ around its central value.  Furthermore, on this domain $r_\pi f_\pi = 0.31(1)$; viz., a constant independent of the current-quark mass, as apparent in Fig.\,\ref{AKFpitimelike}.  Computations show that this remarkable behaviour persists to beyond the charm mass.  

\section{Nucleon}
The nucleon appears as a pole in a six-point quark Green function.  The residue is proportional to the nucleon's Faddeev amplitude, which is obtained from a Poincar\'e covariant Faddeev equation that sums all possible exchanges and interactions which can take place between three dressed-quarks.  A tractable Faddeev equation for baryons was formulated in Ref.\,\cite{Cahill:1988dx}.  It is founded on the observation that an interaction which describes colour-singlet mesons also generates quark-quark (diquark) correlations in the colour-$\bar 3$ (antitriplet) channel \cite{Cahill:1987qr}.  The lightest diquark correlations appear in the $J^P=0^+,1^+$ channels and hence only they are retained in approximating the quark-quark scattering matrix.  While diquarks do not appear in the strong interaction spectrum; e.g., Refs.\,\cite{Bhagwat:2004hn,Bender:1996bb}, the attraction between quarks in this channel justifies a picture of baryons in which two quarks are always correlated as a colour-$\bar 3$ diquark pseudoparticle, and binding is effected by the iterated exchange of roles between the bystander and diquark-participant quarks.  

\begin{figure}[t]
\vspace*{-5ex}

\begin{minipage}[t]{0.45\textwidth}
\leftline{\includegraphics[clip,width=0.9\textwidth]{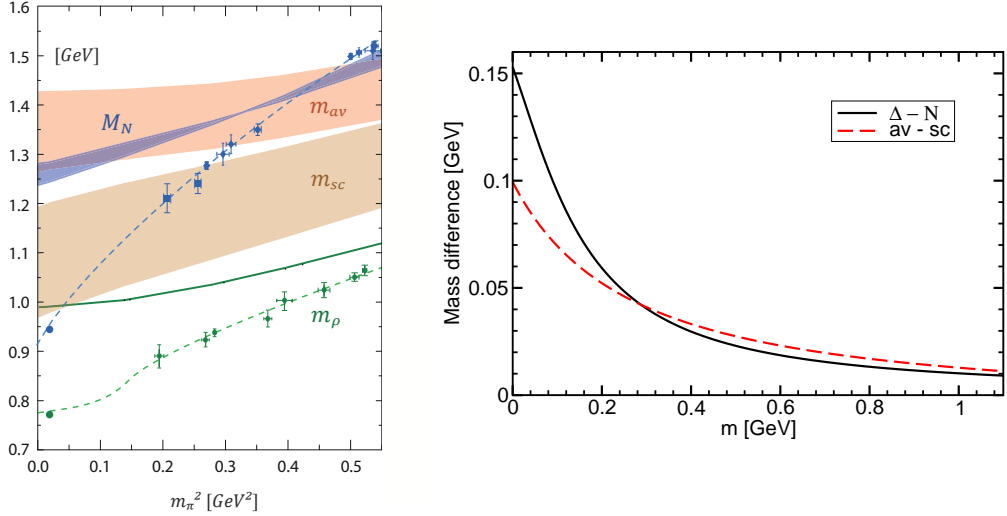}}
\end{minipage}
\begin{minipage}[t]{0.45\textwidth}
\vspace*{-38ex}

\rightline{\includegraphics[clip,width=1.1\textwidth]{FIGS/MassDiff.eps}}
\end{minipage}\vspace*{-1ex}
\caption{\label{nucleonmass} \underline{Left Panel}. 
\emph{Thick bands}: Evolution with current-quark mass, $\hat m$, of the scalar and axial-vector diquark masses: $m_{sc}$ and $m_{av}$.  Bands demarcate sensitivity to the variation in $\omega$.  ($m_\pi$, calculated from rainbow-ladder meson Bethe-Salpeter equation: $\hat m = 6.1\,$MeV $\Rightarrow m_\pi = 0.138\,$GeV.)
\emph{Solid curve}: Evolution of $\rho$-meson mass \protect\cite{Eichmann:2008ae}.  This observable quantity is insensitive to $\omega$.  
With $m_\rho$, results from simulations of lattice-regularised QCD \protect\cite{AliKhan:2001tx} are also depicted along with an analysis and chiral extrapolation \protect\cite{Allton:2005fb}, \emph{short dashed curve}.  
\emph{Thin band}: Evolution with $\hat m$ of the nucleon mass obtained from the Faddeev equation: $\hat m =6.1\,$MeV, $M_N=1.26(2)\,$GeV cf.\ results from lattice-QCD \cite{Ali Khan:2003cu,Frigori:2007wa} and an analysis of such results \cite{Leinweber:2003dg}, \emph{dashed curve}.  (Figure adapted from Ref.\,\cite{Eichmann:2008ef}.) 
\underline{Right Panel}. \emph{Solid curve}, $M_\Delta-M_N$ as a function of current-quark mass, evaluated as described in connection with Eqs.\,(\protect\ref{dqm}) within the framework of Ref.\protect\cite{Cloet:2008wg}; and \emph{dashed curve}, $m_{\rm av}-m_{\rm sc}$.
} 
\end{figure}

Following Refs.\,\cite{Eichmann:2008ae,Cahill:1988dx} one can construct a parameter-free Faddeev equation whose solution describes a nucleon's dressed-quark core.  This enables the simultaneous calculation and unification of meson and nucleon observables within a framework that provides a veracious description of the pion as both a Goldstone mode and a bound state of dressed-quarks.  
The study predicts the evolution of the nucleon mass with a quantity that can methodically be connected with the current-quark mass in QCD.  This is depicted in Fig.\,\ref{nucleonmass}.   Notably, despite the large $\omega$-dependence of the unobservable diquark masses, the nucleon mass is only weakly sensitive to this model parameter.  Again, systematic corrections to the DSE's leading order truncation move results into line with experiment.

It is notable that the $\omega$-band on $m_{av}-m_{sc}$ is much narrower than that on the individual masses, apparent in Fig.\,\ref{nucleonmass}, and that this difference falls with increasing current-quark mass.  
Since the $\Delta$-baryon may only involve axial-vector diquark correlations, the $\Delta$-$N$ mass splitting is correlated with $m_{av}-m_{sc}$.  One can therefore infer that $M_\Delta - M_N$ will depend weakly on $\omega$ and fall with increasing $m_\pi^2$.
Notwithstanding the correlation, near agreement between the experimental value of $M_\Delta - M_N=0.29\,$GeV and $m_{av}-m_{sc} = 0.27(3)\,$GeV at the physical pion mass is incidental \cite{Alkofer:2004yf}.

The relationship between $M_\Delta-M_N$ and $m_{av}-m_{sc}$ can be illustrated using the Faddeev equation model of Ref.\,\cite{Cloet:2008wg}.  The model uses algebraic forms for all elements.  It expresses the evolution of the dressed-quark mass with current-quark mass but not that of the diquark masses, which can reasonably be parametrised based on the following observations.  The mass-splitting is nonzero at the physical light-quark current-mass and yields a particular $\Delta$-nucleon mass splitting.  That value is $0.15\,$GeV in Ref.\,\cite{Cloet:2008wg} but the magnitude is immaterial in what follows. Since spectroscopically relevant corrections to the rainbow-ladder truncation vanish in the heavy$+$heavy-quark limit \cite{Bhagwat:2004hn,Bhagwat:2006xi,Soto:2006zs}, this truncation can be used to determine the qualitative behaviour of $m_{av}-m_{sc}$ with increasing current-quark mass: the difference decreases monotonically to the asymptotic result $m_{av}=m_{sc}$, which is natural because, e.g., in quantum mechanics $m_{av}-m_{sc}$ can only arise through a hyperfine interaction and that vanishes as an inverse power of current-quark mass-squared.  With this motivation the Faddeev equation in Ref.\,\cite{Cloet:2008wg} was solved with
\begin{equation}
\label{dqm}
    m_{sc}(m) = 2 M_Q^E + \frac{0.282}{1 + (M_Q^E/M_u^E)^2}, \;
    m_{av}(m) = 2 M_Q^E + \frac{0.476}{1 + (M_Q^E/M_u^E)^2},
\end{equation}
where $m$ is the current-quark mass and the Euclidean constituent-quark mass is defined via 
$M_Q^E(m)= \{ p \,|\, p^2 = M^2(p^2,m), p>0\}.$
(NB.\ With increasing current-quark mass $M_Q^E(m) - m \to 0^+\!$.)

The result is depicted in the right panel of Fig.\,\ref{nucleonmass}.  From this figure and studies underway one can make the following observations, which are general features of Faddeev equations in the class under consideration.  $M_\Delta-M_N > m_{av}-m_{sc}$ in the chiral limit and at the physical light-quark current-mass.  This is consistent with a further reduction in $M_\Delta-M_N$ owing to the so-called pseudoscalar meson cloud.  Such contributions vanish with increasing current-quark mass so that the quark core \emph{becomes} the baryon.  At a particular current-quark mass, which depends, e.g., on the model's chiral limit value of $M_\Delta-M_N$, $m_{av}-m_{sc}$ becomes greater than $M_\Delta-M_N$.  This remains true thereafter.  With increasing current-quark mass $M_\Delta-M_N \to 0^+$.  Finally, $M_\Delta(m)/[M_Q^E(m)+m_{av}(m)]\to 1^+$ as $m \to \infty$.  (NB.\ $M_\Delta(m)=M_N$ and $m_{av}=m_{sc}$ in this limit.)  

In order to calculate nucleon form factors the Faddeev equation must be augmented with a nucleon-photon current that automatically preserves the Ward-Takahashi identity for on-shell nucleons described by the Faddeev amplitude \cite{Oettel:1999gc}.  Following this one can produce nucleon form factors with realistic $Q^2$-evolution \cite{Eichmann:2008ef}.  A notable prediction is $r_1^{nu} > r_1^{nd}$; viz., that the Dirac radius of the $u$-quark in the neutron is larger than that of the $d$-quark.  This result is consistent with contemporary parametrisations of experimental data and owes to the presence of axial-vector diquark correlations in the nucleon.  

\section{Strangeness}
The role played by $s$-quarks in light-hadron structure has long been of interest.  Their contribution to nucleon form factors is accessible via parity violating electron-proton scattering \cite{Arrington:2006zm,Beise:2004py,Young:2006jc}.  The natural magnitude of the contribution may be estimated by considering the $s$-quark content of a dressed $u$-quark.  The gluon vacuum polarisation appears in the gap equation's kernel.  It includes $u$-, $d$- and $s$-quark contributions.  Since the $g \to \bar q q$ vertex is flavour-independent and the polarisation diagram contains two quark propagators, then in perturbation theory the infrared behaviour of the vacuum polarisation is regularised by the current-quark mass and receives a contribution related to $1/m_f^2$, where $m_f$ is the current-quark mass of flavour $f$.  If one defines the $u$-quark content of the vacuum polarisation to be $\Pi_u$, then based on contemporary estimates of the current-quark masses 
\begin{equation}
\Pi_d = \frac{m_u^2}{m_d^2}\, \Pi_u = 0.2 \, \Pi_u\,, \;
\Pi_s = \frac{m_d^2}{m_s^2}\, \Pi_d = 0.003 \, \Pi_d = \frac{m_u^2}{m_s^2}\, \Pi_u = 0.0005 \, \Pi_u\,.
\end{equation}
Thus from perturbation theory one does not expect a noticeable $s$-quark content in the dressed $u$- and $d$-quarks and also therefore not in the nucleon.  It is nonetheless conceivable that nonperturbatively the result is otherwise.

The flavour content of the gluon vacuum polarisation is not active in the rainbow truncation of the gap equation.  Consider, however, a vertex correction wherein: the dressed-quark emits a gluon; that gluon splits into $\bar q q$; one of these fermions emits a gluon; that gluon is absorbed on the through-propagating dressed-quark line; the $\bar q q$ then proceed to recombine as a gluon; that gluon is finally absorbed by the through-propagating dressed-quark line.  With this type of vertex correction, the through-propagating dressed-quark interferes with the quarks in the gluon vacuum polarisation.  Naturally, the $\bar q q$ intermediate state could emit any number of gluons that are absorbed by the through-propagating dressed-quark line.  This consideration shows that vertex corrections generate a resonant (meson$+$quark- or diquark$+$antiquark-loop) contribution to the gap equation and a continuum (non-resonant) contribution.  In the following the meson contribution is estimated.  The nonresonant contribution should be of the same order as the perturbative result.  Furthermore, since the mass-squared of a $us$-diquark is more than four-times that of a kaon \cite{Burden:1996nh}, diquark contributions should be materially suppressed relative to those from kaon-like correlations.

These considerations lead to the following correlation-augmented rainbow gap equations, in which are assumed $m_u=m_d$ and a mass-independent renormalisation scheme:
\begin{eqnarray}
\nonumber 
\lefteqn{
S_u^{-1}(p) = Z_2(i \gamma\cdot p + m^{\rm bm}_u) +  Z_1 \int^\Lambda_q\! g^2 D_{\mu\nu}(p-q) \frac{\lambda^a}{2}\gamma_\mu S_u(q) \frac{\lambda^a}{2}\gamma_\nu} \\
\nonumber
&+ &  3 \left(\frac{1}{2M^D_u}\right)^2 \int\frac{d^4 q}{(2\pi)^4} \, \Delta_{\pi}(q) \bar\Gamma_\pi(p+q/2;-q)\gamma\cdot q \, S_u(p-q) \, \Gamma_\pi(p+q/2;q)\gamma\cdot q\\
&+ &  2 \left(\frac{1}{2M^D_s}\right)^2 \int\frac{d^4 q}{(2\pi)^4} \, \Delta_{K}(q) \bar\Gamma_K(p+q/2;-q)\gamma\cdot q \, S_s(p-q) \, \Gamma_K(p+q/2;q)\gamma\cdot q\,,\\
\nonumber 
\lefteqn{
S_s^{-1}(p) = Z_2(i \gamma\cdot p + m^{\rm bm}_u) +  Z_1 \int^\Lambda_q\! g^2 D_{\mu\nu}(p-q) \frac{\lambda^a}{2}\gamma_\mu S_s(q) \frac{\lambda^a}{2}\gamma_\nu} \\
&+ &  4 \left(\frac{1}{2M^D_s}\right)^2 \int\frac{d^4 q}{(2\pi)^4} \, \Delta_{K}(q) \bar\Gamma_K(p+q/2;-q)\gamma\cdot q \, S_s(p-q) \, \Gamma_K(p+q/2;q)\gamma\cdot q\,.
\end{eqnarray}
In these equations: $M^D_q$ is the dynamical constituent-quark mass, defined as the dressed-quark mass function evaluated at the origin in momentum space; $\Delta_M(q)$ is a free-particle pseudoscalar meson propagator; $\Gamma_M$ is the Bethe-Salpeter amplitude for the associated meson; and the $\gamma\cdot q$ factors enforce a pseudovector coupling between mesons and a dressed-quark.

To obtain a first estimate of the magnitude of the $s$-quark content, one can use a NJL-like model, defined as follows.  The renormalisation constants in the gap equation are set equal to one; the rainbow part is expressed via
\begin{equation}
Z_1 \int^\Lambda_q\! g^2 D_{\mu\nu}(p-q) \frac{\lambda^a}{2}\gamma_\mu S_f(q) \frac{\lambda^a}{2}\gamma_\nu  \rightarrow  \frac{1}{m_G^2}  \int \frac{d^4 q}{(2\pi)^4}\, \theta(\Lambda^2-q^2) \frac{\lambda^a}{2}\gamma_\mu S_f(q) \frac{\lambda^a}{2}\gamma_\mu \,;
\end{equation}
and the resonant vertex correction contribution through
\begin{eqnarray}
\nonumber\lefteqn{\left(\frac{1}{2M^D_f}\right)^2 \int\frac{d^4 q}{(2\pi)^4} \, \Delta_{M}(q) \bar\Gamma_M(p+q/2;-q)\gamma\cdot q \, S_f(p-q) \, \Gamma_M(p+q/2;q)\gamma\cdot q }\\
&
\rightarrow &
\left(\frac{1}{2 f_M^2}\right)^2 \int\frac{d^4 q}{(2\pi)^4} \, \theta(\Lambda_f^2-q^2) \, \Delta_{M}(q) \gamma_5\gamma\cdot q \, S_f(p-q) \, \gamma_5\gamma\cdot q\,.
\label{mesongap}
\end{eqnarray}
The parameters are: $m_G$, a gluon mass-scale; $\Lambda$, a NJL cutoff; and $\Lambda_f$, which are cutoffs for each quark flavour owing to finite meson size as expressed through the Bethe-Salpeter amplitudes.  In this estimate experimental values of the leptonic decay constants are used.  Naturally, in a detailed calculation they would be calculated quantities.  Since only the gluon provides interaction strength in the far ultraviolet, then $\Lambda > \Lambda_f$.  Moreover, the dominant piece of the kaon Bethe-Salpeter amplitude drops off faster than the kindred pion amplitude \cite{Maris:1997tm}, hence $\Lambda_u \geq \Lambda_s$.

Owing to the quark propagator, Eq.\,(\ref{mesongap}) involves a single angular integral, which means the right-hand-side depends on $p^2$.  However, in this first estimate it is expedient to approximate $A_f$, $B_f$ as momentum-independent and therefore introduce a mean-square average $\bar p_f^2$, defined via
\begin{equation}
\bar p_f^2 \int_0^{\Lambda_f^2} ds\, \frac{s}{s+M_f^2}  := \int_0^{\Lambda_f^2} ds\, \frac{s^2}{s+M_f^2}\,,
\end{equation}
to be used wherever $p^2$ appears.  It has been verified that the results are not sensitive to the value of  $\bar p_f^2$.  NB.\ $M_f$ is the dressed $f$-quark mass.  

To complete the estimate the following parameter values are chosen: $m_G^2 = [ 0.55/(3 \pi^2)] \Lambda^2$, $\Lambda = 1\,$GeV; $f_\pi = 0.092\,$GeV, $f_K=0.11\,$GeV; and $\Lambda_u = 0.8\,\Lambda$.  Setting $\Lambda_s = \Lambda_u$ maximises the model's achievable $s$-quark content.

In the case $m_u=0$, $m_s=0.12\,$GeV and with no coupling to meson loops the reference rainbow-truncation result is obtained:
\begin{equation}
A_u = 1\,, A_s = 1\,, M_u = 0.56\,{\rm GeV}, M_s = 0.70\,{\rm GeV}, \langle \bar q q \rangle = (-0.29\,{\rm GeV})^3.
\end{equation}
Adding only the pion loop these results become
\begin{equation}
A_u = 1.12\,, A_s = 1\,, M_u = 0.38\,{\rm GeV}, M_s = 0.70\,{\rm GeV}, \langle \bar q q \rangle = (-0.26\,{\rm GeV})^3.
\end{equation}
Evidently, in the absence of $s$-quarks the dressed-u quark contains 7\% $\pi^+$ and 4\% $\pi^0$; i.e., the probability of finding a dressed $d$-quark in the dressed $u$-quark is 7\%.  (NB.\ These probabilities are read from the wave function renormalisation; viz., $1/A_f$.)   Including the kaon loop, too:
\begin{equation}
A_u = 1.16\,, A_s = 1.07\,, M_u = 0.32\,{\rm GeV}, M_s = 0.63\,{\rm GeV}, \langle \bar q q \rangle = (-0.25\,{\rm GeV})^3,
\label{chiralcorrection}
\end{equation}
from which one reads that there is 3\% chance of finding an $s$-quark in a dressed $u$-quark and a 7\% chance of finding a $u$- or $d$-quark in a dressed $s$-quark.  Equation~(\ref{chiralcorrection}) confirms the magnitudes of these corrections assumed in Ref.\,\cite{Eichmann:2008ae}.

In the case of physical light-quark current-masses; viz., $m_u=0.005\,$GeV, $m_s=0.12\,$GeV:
\begin{equation}
A_u = 1.15\,, A_s = 1.07\,, M_u = 0.33\,{\rm GeV}, M_s = 0.63\,{\rm GeV}.
\end{equation}
This reveals a 2\% chance of finding an $s$-quark in a dressed $u$- or dressed $d$-quark, and a 7\% chance of finding a $u$- or $d$-quark in a dressed $s$-quark.  Implemented in an independent-particle constituent-quark-like model (e.g., App.~D, Ref.\,\cite{Cloet:2008wg}) this corresponds to $\mu_p^S \approx - 0.02\,$nuclear magnetons.  

\begin{figure}[t]
\rightline{\includegraphics[clip,width=0.47\textwidth]{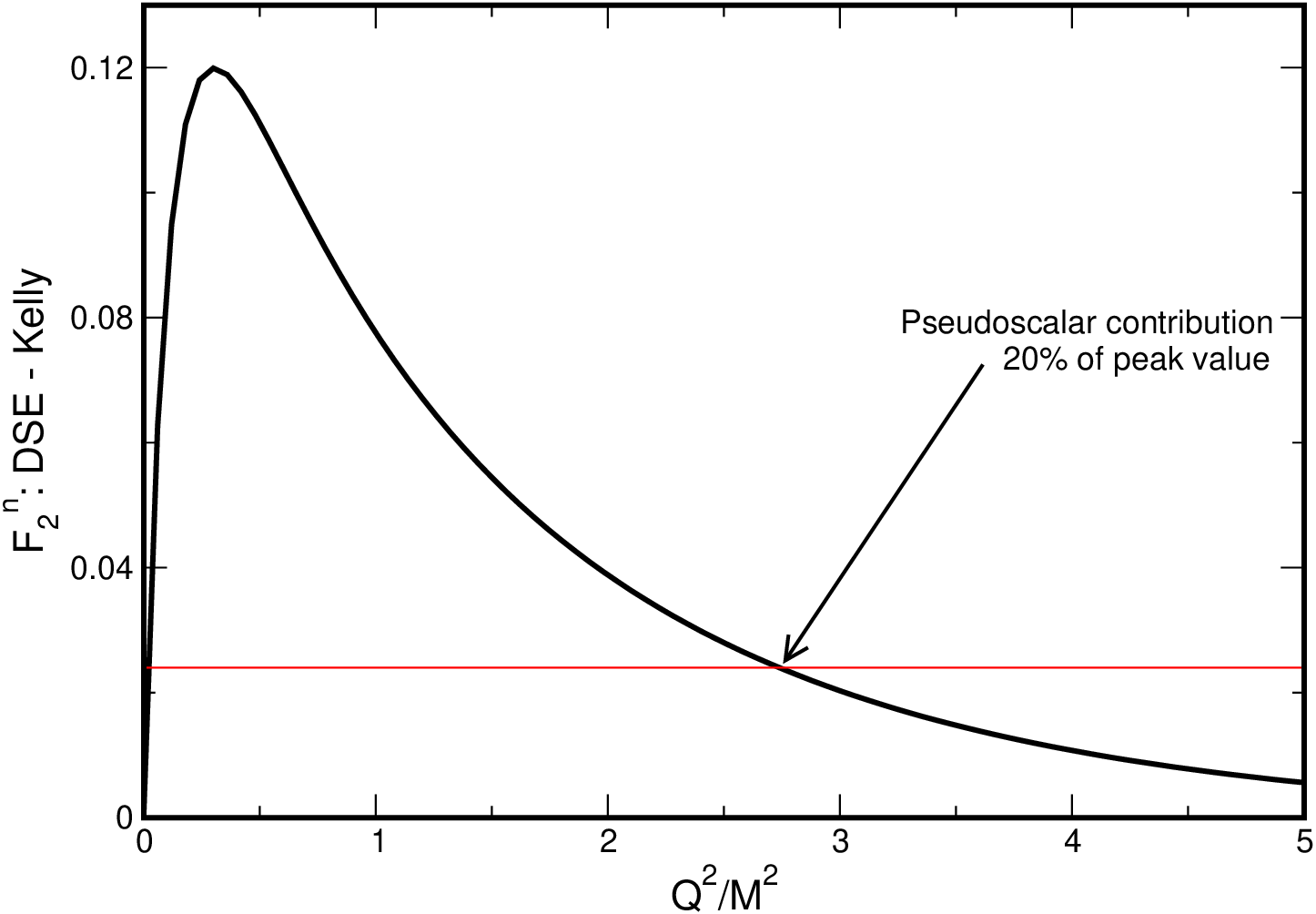}}
\vspace*{-29.5ex}

\parbox{0.5\textwidth}{%
\caption{\label{picloud} Difference between a parametrisation \cite{Kelly:2004hm} of experimental results relating to the neutron's Pauli form factor, $F_2^n$, and a computation based on Refs.\protect\cite{Alkofer:2004yf,Cloet:2008wg}.  The latter expresses the result from an impulse-like approximation expressed in terms of dressed-quarks, which explicitly omits contributions from the pseudoscalar meson cloud.  By $Q^2=3\,M_N^2$ the difference has fallen to $<20$\% of its peak value, a result suggestive of the quark-core achieving primacy at this point.}}
\end{figure}

\section{Epilogue}
Dynamical chiral symmetry breaking exists in QCD.  It is manifest in dressed-quark and -gluon propagators, and in dressed vertices.  DCSB predicts, amongst other things, that the light-quark mass function becomes massive at infrared momenta, and that pseudoscalar mesons are remarkably light and couple very strongly to the lightest baryons.  DCSB means that the Higgs mechanism is largely irrelevant to the bulk of normal matter in the universe.  
 
Form factors are a primary means by which to explore and chart the structure of hadrons. One may anticipate that the near- to medium-term will see progress in quantifying effects owing to the pseudoscalar meson cloud (see Fig.\,\ref{picloud}), locating the transition from the nonperturbative to the perturbative domain within QCD, elucidating the connection between the spacelike and timelike behaviour of form factors, and explaining the relationship between parton properties on the light-front and the rest frame structure of hadrons.

\medskip

\hspace*{-\parindent}\textbf{Acknowledgments.}~~~Conversations and correspondence with 
G.~Eichmann, B.~El-Bennich, R.\,J.~Holt, T.~Kl\"ahn, A.~Krassnigg, D.~Nicmorus, R.\,D.~Young and J.M.~Zanotti are acknowledged; as are:
Department of Energy, Office of Nuclear Physics, contract nos.\ DE-FG03-97ER4014 and DE-AC02-06CH11357; 
and use of ANL's Computing Resource Center's facilities.

\end{document}